\documentclass[11pt,aps,prb,twocolumn,reprint]{revtex4-1}
\usepackage{graphicx}
\usepackage{amsmath}

\begin{document}
\title{Voltage-controlled surface plasmon-polaritons in double graphene layer structures}

\author{D. Svintsov}
\affiliation
{Institute of Physics and Technology, Russian Academy of Sciences, Moscow 117218, Russia}

\author{V. Vyurkov}
\affiliation
{Institute of Physics and Technology, Russian Academy of Sciences, Moscow 117218, Russia}

\author{V. Ryzhii}
\affiliation
{Research Institute for Electrical Communication, Tohoku University, Sendai 980-8577,  Japan}
\affiliation
{Japan Science and Technology Agency, CREST, Tokyo 107-0075, Japan}

\author{T. Otsuji}
\affiliation
{Research Institute for Electrical Communication, Tohoku University, Sendai 980-8577, Japan}
\affiliation
{Japan Science and Technology Agency, CREST, Tokyo 107-0075, Japan}

\begin{abstract}
We study the spectra and damping of surface plasmon-polaritons in double graphene layer structures. It is shown that application of bias voltage between layers shifts the edge of plasmon absorption associated with the interband transitions. This effect could be used in efficient plasmonic modulators. We reveal the influence of spatial dispersion of conductivity on plasmonic spectra and show that it results in the shift of cutoff frequency to the higher values.
\end{abstract}

\maketitle

\section{Introduction}
Recently the possibility of fabrication of double-layer graphene structures was reported\cite{Role_of_graphene}. In such structures two graphene layers separated by thin dielectric are clad between the two sections of substrate. A number of devices based on double-layer graphene has already been fabricated, among them there are field-effect tunneling transistors~\cite{Britnell} and optical modulators~\cite{Double-layer-modulator}. Plasmonic properties of double-layer graphene are of particular interest too: first, such structures can act as plasmonic modulators~\cite{Andersen}; second, plasma resonances can affect the characteristics of optical modulators, resulting in greater modulation depth near plasmon peak~\cite{Modulator-concept-analysis}. However, the spectra of plasma oscillations in double graphene in the presence of bias voltage between layers have not been studied yet.

In this paper we derive the spectra of surface plasmon-polaritons (SPP) in double-graphene layer (2GL) structures using the model of graphene conductivity presented in~\cite{Falkovsky-Varlamov}. In our calculations the spatial dispersion of conductivity is taken into account, which turns out of great importance at the elevated Fermi energy $\mu$ comparable of higher than the temperature $T$. We analyze the effect of bias voltage applied between layers on the spectra and damping of SPP and show that the 2GL structures can be used as efficient plasmonic modulators.

\section{Electrodynamics of surface plasmon-polaritons in 2GL}
\begin{figure}
\includegraphics[width=1\linewidth]{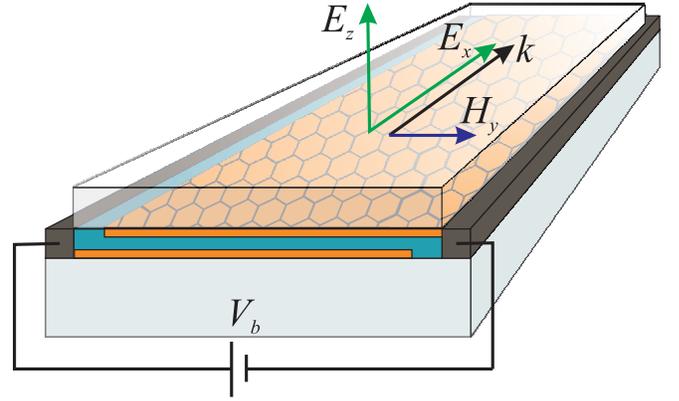}
\caption{\label{Schematic} Schematic view of the 2GL structure and directions of vectors ${\bf E}$, ${\bf H}$ and ${\bf k}$ in the TM SPP}
\end{figure}

The 2GL structure under consideration is schematically shown in Fig.~\ref{Schematic}. It consists of two GLs separated by a barrier layer. One of the edges of each GL is connected with a contact, while the other edge is isolated. The bias voltage $V_b$ is applied between the contacts. Thus, one GL serves as the gate for another. The dielectric constants of the top, middle, and bottom media are denoted by $\varepsilon_t$, $\varepsilon_m$, and $\varepsilon_b$, respectively; the distance between GLs is $d$. The $x$-axis coincides with the direction of the SPP propagation parallel to the metal electrodes, the $z$-axis is orthogonal to the graphene plane. 

Due to the peculiarities of graphene conductivity, the propagation of TE SPP modes along its surface is prohibited~\cite{Hanson}. For this reason we focus on TM-modes with the ac electric and magnetic fields ${\bf E}=\{E_x,\, 0,\, E_z\}$, ${\bf H}=\{0,\, H_y,\, 0\}$. We search for the solutions of Maxwell equations in the form:
\begin{gather}
{\bf E}_t={\bf E}_{0\,t} e^{i (k x -\omega t)-\kappa_t z},\\
{\bf E}_m=\left({\bf E}_{0\,m \uparrow } e^{\kappa_m z}+{\bf E}_{0\,m \downarrow } e^{-\kappa_m z}\right) e^{i (k x -\omega t)},\\
{\bf E}_b={\bf E}_{0\,b} e^{i (k x -\omega t)+\kappa_b z},
\end{gather}
where the indices $t$, $m$, and $b$, again, stand for the top, middle and bottom media, respectively, $\kappa_i=\sqrt{k^2-\varepsilon_i\omega^2/c^2}$, and $c$ is the speed of light.

Matching the vectors $\bf E$ and $\bf H$ on the graphene surfaces (the conductivities of top and bottom layers are $\sigma_t$ and $\sigma_b$), one arrives at the dispersion law
\begin{multline}
\label{General}
\left[ \frac{i\omega}{4\pi}\left( \frac{\varepsilon_t}{\kappa_t}+\frac{\varepsilon_m}{\kappa_m}\right)-\sigma_t \right] \left[ \frac{i\omega}{4\pi} \left(\frac{\varepsilon_b}{\kappa_b} +\frac{\varepsilon_m}{\kappa_m}\right) - \sigma_b \right]{{e}^{2\kappa_m d}} =\\
\left[\frac{i\omega}{4\pi} \left(\frac{\varepsilon_b}{\kappa_b} - \frac{\varepsilon_m}{\kappa_m}\right)- \sigma_b \right]
\left[\frac{i\omega}{4\pi} \left(\frac{\varepsilon_t}{\kappa_t} - \frac{\varepsilon_m}{\kappa_m}\right)- \sigma_t \right].
\end{multline}
Eq.~(\ref{General}) is simplified for symmetric structures with $\varepsilon_t=\varepsilon_b=\varepsilon$ and $\sigma_t=\sigma_b=\sigma$. The latter condition is valid provided the residual doping of graphene layers is weak. In the situation at any finite bias voltage the modules of Fermi energies in GLs are equal but have opposite signs, hence, due to electron-hole symmetry the conductivities of layers are equal. The simplified dispersion laws are
\begin{gather}
\label{symmetric}
\left( \frac{i\omega \varepsilon }{\kappa}-4\pi \sigma \right)+\frac{i\omega \varepsilon_m}{\kappa_m}\tanh \left( \frac{\kappa_m d}{2} \right)=0,\\
\label{antisymmetric}
\left( \frac{i\omega \varepsilon }{\kappa}-4\pi \sigma \right)\tanh \left( \frac{\kappa_m d}{2} \right)+\frac{i\omega \varepsilon_m}{\kappa_m}=0.
\end{gather}
Eq.~(\ref{symmetric}) corresponds to a "symmetric mode" with the current oscillations in both layers in phase, while Eq.~(\ref{antisymmetric}) represents an "antisymmetric mode". We are particularly interested in symmetric oscillations~\cite{Asymmetric-Reference}. At small distances between layers ($\kappa_m d \ll 1$) their dispersion depends only on the net conductivity of two layers~\cite{Dubinov-Aleshkin}:
\begin{equation}
\frac{\varepsilon_t}{\kappa_t}+\frac{\varepsilon_b}{\kappa_b}=-\frac{4 \pi i}{\omega}\times 2\sigma.
\end{equation}

\section{Model of graphene conductivity}
To perform a numerical analysis of SPP spectra we use the model of graphene conductivity, presented in Ref.~\cite{Falkovsky-Varlamov}:
\begin{widetext}
\begin{multline}
\label{VV-Varlamov}
{{\sigma }_{\bf{k}\omega }}=\frac{i{e^2}}{\hbar {\pi^{2}}}\sum\limits_{a=1,2}{\int{\frac{{d^2}{\bf p}v_{x}^{2}\left\{ f\left[ {{\epsilon }_{a}}\left( {{\bf p}_-} \right) \right]-f\left[ {{\epsilon }_{a}}\left( {{\bf{p}}_{+}} \right) \right] \right\}}{\left[ {\epsilon_a}\left( {{\bf p}_+} \right)-{\epsilon_a}\left( {{\bf{p}}_{-}} \right) \right]\left[ \hbar \omega -{\epsilon_a}\left( {{\bf p}_+} \right)+{\epsilon_a}\left( {{\bf{p}}_{-}} \right) \right]}}}+\\
\frac{2i{e^2}\hbar \omega }{\hbar^2 \pi^2}\int{\frac{{d^2}{\bf p}{v_{21}}{v_{12}}\left\{ f\left[ {\epsilon_1}\left( {{\bf{p}}_{-}} \right) \right]-f\left[ \epsilon_2\left( {{\bf{p}}_{+}} \right) \right] \right\}}{\left[ \epsilon_2\left( {{\bf{p}}_{+}} \right)-{\epsilon_1}\left( {{\bf{p}}_{-}} \right) \right]\left[ {{\left( \hbar \omega  \right)}^{2}}-\left[ \epsilon_2\left( {{\bf{p}}_+} \right)-\epsilon_1\left( {{\bf p}_-} \right) \right]^2 \right]}}.
\end{multline}
\end{widetext}
Here the indices $1$ and $2$ stand for conductance and valence bands, in particular $\epsilon_1({\bf p})=pv_F$ and $\epsilon_2({\bf p})=-pv_F$, $v_F\simeq 10^6$ m/s, ${\bf p}_{\pm}={\bf p}\pm \hbar {\bf k}/2$, $f(\epsilon)$ is the electron distribution function (the equilibrium Fermi function is assumed), $v_x=v_F \cos \theta_{\bf p}$ and $v_{12}=i v_F \sin\theta_{\bf p}$ are the matrix elements of velocity operator. The first line of Eq.~(\ref{VV-Varlamov}) corresponds to the intraband transitions, while the second one accounts for the intraband. To account for the momentum relaxation the frequency should be treated as $ \omega \rightarrow \omega+i\tau_p^{-1}$. The expressions for $\tau_p^{-1}$ for different scattering mechanisms could be found in~\cite{Transport-review}. 
Being interested in high-quality samples with $\mu\gg T$ we restrict ourselves to the consideration of electron-phonon scattering. The appropriate relaxation time is~\cite{Vasko-Ryzhii}
$$
\tau_{p}^{-1}=\tau_0^{-1}\frac{pv_F}{T},
$$
where $\tau_0^{-1}\simeq 3 \cdot 10^{11}$ is the characteristic electron-phonon collision frequency at room temperature $T=26$ meV.

We stress that in the following calculations the spatial dispersion of conductivity could be neither omitted~\cite{Dubinov-Aleshkin} nor treated in a perturbative manner~\cite{Bujan}. At elevated Fermi energies and at high dielectric constants of surrounding materials the dispersion $k(\omega)$ of SPP is close to the threshold line $k=\omega/v_F$, hence, the electronic transitions are substantially indirect. The latter leads to a pronounced reduction in interband absorption compared to the direct case (the imaginary part of conductivity chanes too). To clarify the effect formally we pass to the elliptic coordinates in Eq.~(\ref{VV-Varlamov}): $2p_x=k\cosh u \cos v$, $2p_y=k \sinh u\sin v$. Note that the wave vector is complex and the functions $|{\bf p}\pm \hbar {\bf k}/2|$ should be treated as absolute values of complex vectors. Extracting the real part of conductivity with the Sokhotski theorem in the limit $\tau^{-1}\rightarrow 0$ we arrive at
\begin{widetext}
\begin{gather}
\label{Inter-spatial}
\operatorname{Re}\sigma_{inter} =\frac{e^2}{2\pi\hbar} \sqrt{1-{q^2}} \int\limits_0^\pi {\left\{ f\left[ -\frac{\hbar \omega}{2}(1+q\cos v) \right]-f\left[ \frac{\hbar \omega}{2}(1-q\cos v) \right] \right\}\frac{1-q^2 \cos^2 v}{1-q^2 \sin^2 v} \sin^2 vdv},\\
\label{Intra-spatial}
\operatorname{Re}\sigma_{intra}=\frac{e^2}{2\pi\hbar }{{\left[ q^2-1 \right]}^{-1/2}}\int\limits_{0}^{\infty }{\left\{ f\left[ \frac{\hbar \omega }{2}(q\cosh u-1) \right]-f\left[ \frac{\hbar\omega}{2}(q\cosh u+1) \right] \right\}\frac{q^2 \cosh^2u-1}{q^2 \sinh ^2u - 1} \cosh^2udu},
\end{gather}
\end{widetext}
where the dimensionless measure of spatial dispersion $q=|k|v_F/\omega$ is introduced.

\section{Results and discussion}
\begin{figure}
\includegraphics[width=0.95\linewidth]{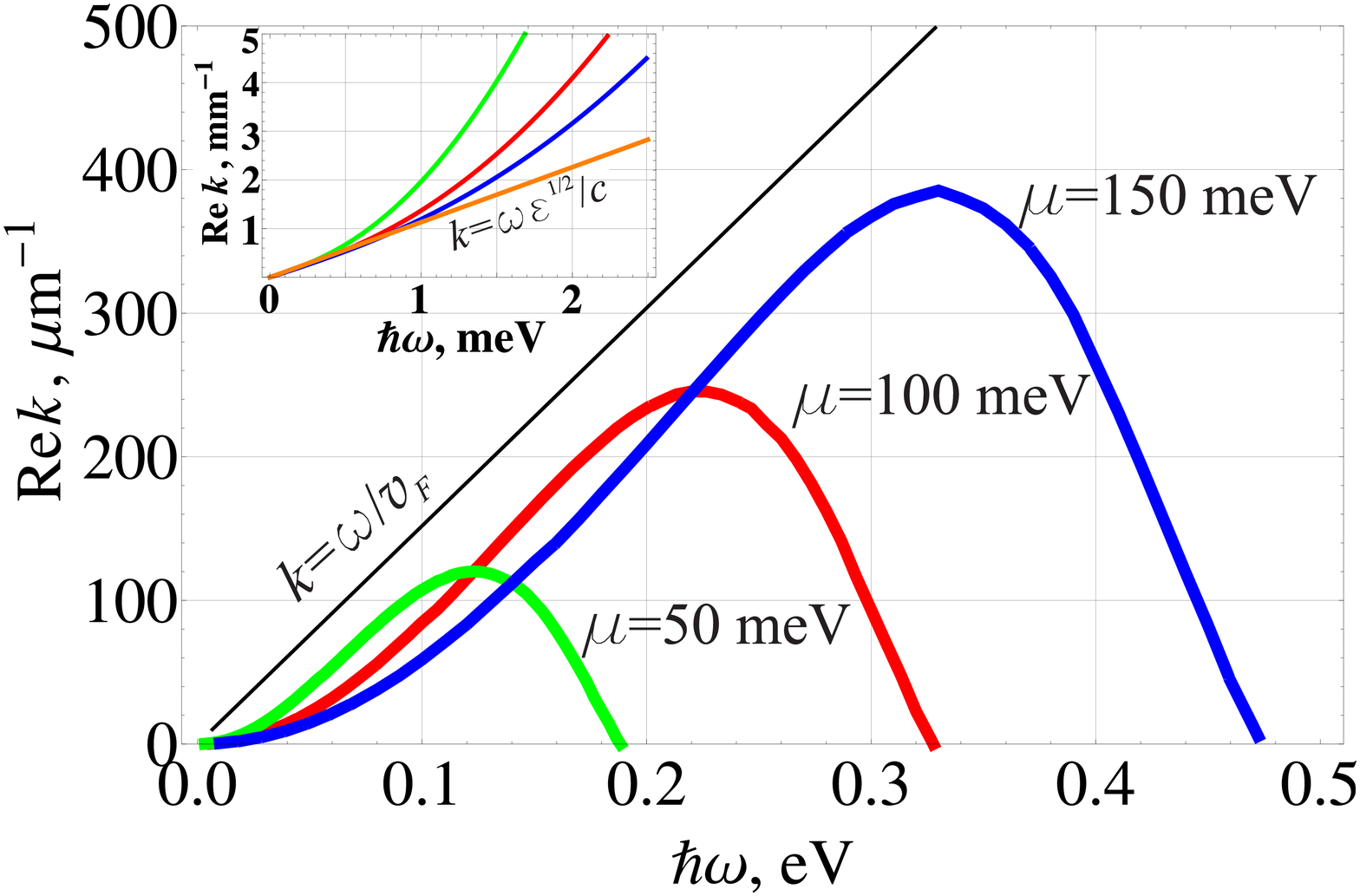}
\caption{\label{Sym} Dispersions $\operatorname{Re}k(\omega)$ of SPP in 2GL at different Fermi energies. Inset: low-energy ''photonic'' behavior of spectra}
\end{figure}

\begin{figure}
\includegraphics[width=0.95\linewidth]{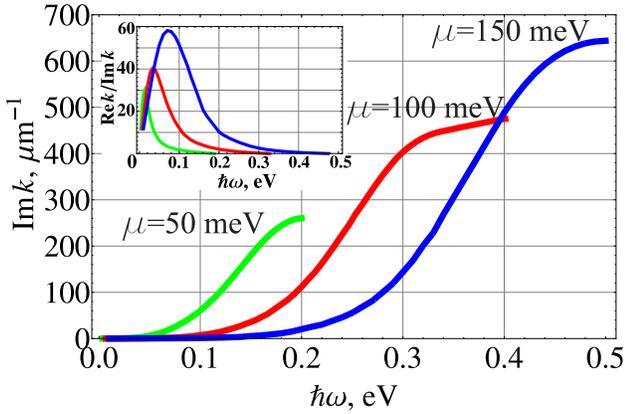}
\caption{\label{ImSym} Damping $\operatorname{Im} k(\omega)$ of SPP in 2GL at different Fermi energies. Inset: ''quality factor'' of SPP defined as $\operatorname{Re} k/\operatorname{Im} k$ at the same Fermi energies}
\end{figure}

The spectra $\operatorname{Re} k(\omega)$ and $\operatorname{Im} k(\omega)$ of SPP calculated with Eqs.~(\ref{symmetric}) and (\ref{VV-Varlamov}) are presented in Figs.~\ref{Sym} and~\ref{ImSym}. The structure parameters are $\varepsilon_t=\varepsilon_m=\varepsilon_b=5$ (boron nitride~\cite{HBN-dielectric}), $d=2$ nm. The three characteristic regions can be singled out in the spectra.

{\it Small wave vectors and frequencies} $k \ll 8 (e^2/\hbar c)\mu/(\hbar c)$: the dispersion is purely ''photonic'', i.e. the dispersion law reads
\begin{equation}
k_x=\frac{\omega \sqrt{\varepsilon}}{c}.
\end{equation}

{\it Intermediate wave vectors and frequencies} $8 (e^2/\hbar c)\mu/(\hbar c) \ll k \ll \omega/v_F$, $\omega \ll 2\mu/\hbar$. In this situation the coupling with photons is negligible and the spatial dispersion of conductivity is negligible too. For this reason we can set $\kappa \approx k$ and obtain the SPP dispersion in the form
\begin{equation}
\label{Simple-dispersion}
k=\frac{i\omega\varepsilon}{8\pi\sigma_{{\bf k}=0,\omega}},
\end{equation}
which yields a well-known square dependence~\cite{Plasmons-RPA} $\operatorname{Re} k \propto \omega^2$ until the interband transitions start playing significant role. 

{\it High frequencies} $\omega \sim 2 \mu/\hbar$. A simple analysis based on Pauli exclusion principle states that the direct interband transitions are allowed at frequencies $\omega>2\mu/\hbar$. At such frequencies the incident light is effectively absorbed, which governs the operation of graphene-based optical modulators~\cite{Double-layer-modulator}. The SPP at such high frequencies are also expected to undergo strong damping. However, the situation for SPP is different as the interband transitions in the case are substantially indirect. From Eq.~(\ref{Inter-spatial}) it is readily seen that the interband absorption at finite ${\bf k}$ is not as strong as at ${\bf k}=0$; in the threshold case $|{\bf k}|>\omega/v_F$ the interband transitions are forbidden at all due to nonconservation of energy and momentum. On the contrary, intraband transitions at $|{\bf k}|>\omega/v_F$ are extremely strong as $\operatorname{Re} \sigma_{intra}$ exhibits a singular behavior $\operatorname{Re} \sigma_{intra}\propto[(kv_F/\omega)^2-1]^{-1/2}$.

\begin{figure}
\includegraphics[width=0.95\linewidth]{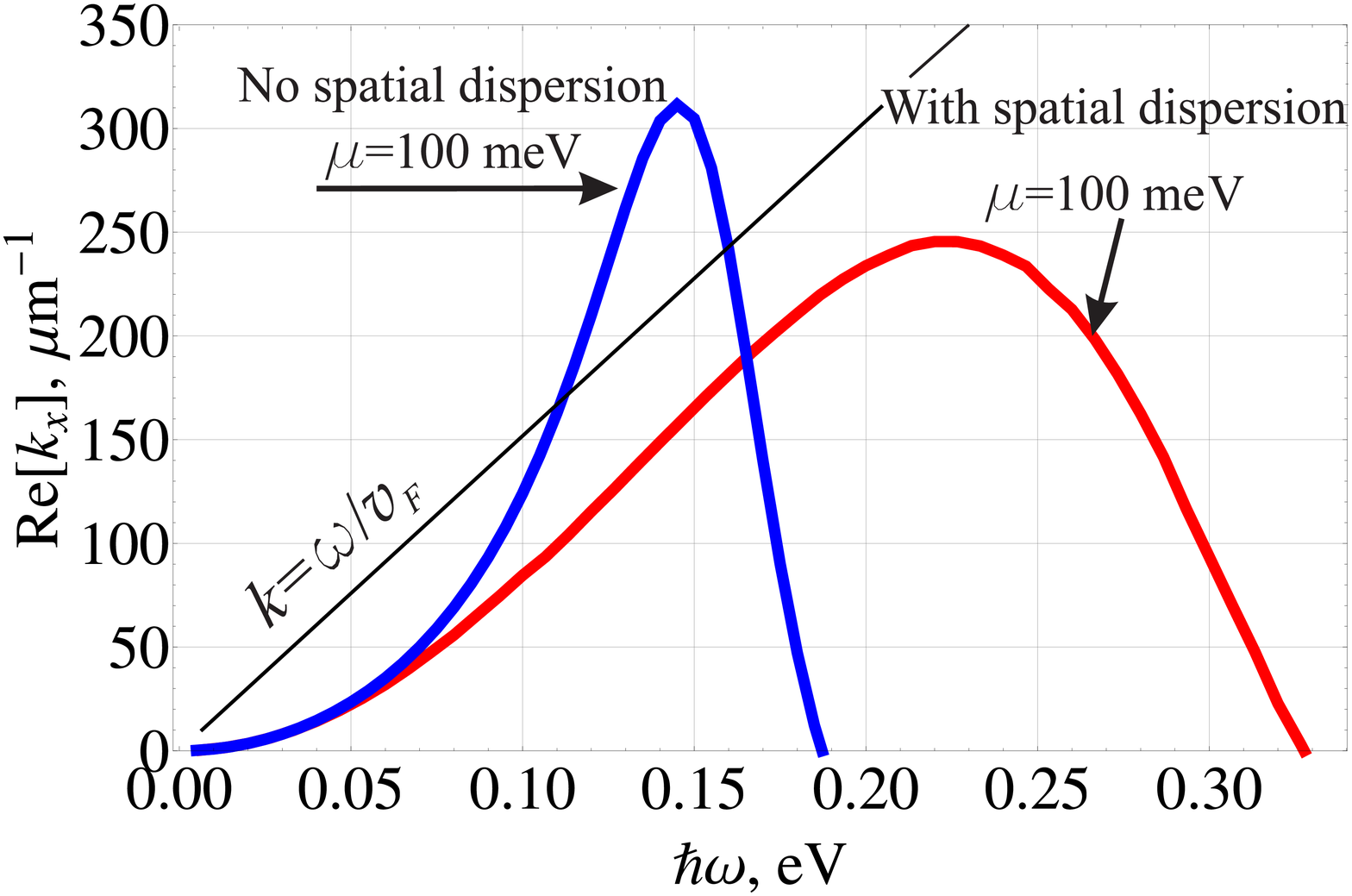}
\caption{\label{Comp} Comparison of SPP dispersions obtained with and without accounting for the spatial dispersion of conductivity}
\end{figure}

The lack of interband transitions at finite wave vectors leads to the higher cutoff frequencies of SPP in graphene compared to photons. The situation is illustrated in Fig.~\ref{Comp}, where the two approaches for the calculation of plasmonic spectra are compared. The dashed line is plotted without accounting for the spatial dispersion of conductivity, while in the solid curve it is taken into account. It should be empathized that even the {\it intrinsic} spatial dispersion due to nonzero wave vector of SPP significantly affects the spectra. An {\it extrinsic} spatial dispersion of conductivity (e.g. created by spatially periodic gates~\cite{Polaritonic-crystal} or surface roughness~\cite{Complete-absorption}) can also change the characteristics of SPP cardinally. A simple analysis of relation~(\ref{Simple-dispersion}) yields that the higher the permittivity $\varepsilon$ and the Fermi energy $\mu$ are, the more important the intrinsic spatial dispersion is.

Despite the shift of SPP cutoff frequency to higher values, the modulation of propagation length by applying bias voltage $V_b$ between layers is still efficient. To study this dependence we implement the plate-capacitor model relating $V_b$ to the charge density $\rho$ in layers
\begin{equation}
\label{Capacitor}
\frac{\varepsilon_m}{4\pi d}(V_b-2\mu)=\rho;
\end{equation}
positive voltage $V_b$ corresponds to the positive charge density. In the limit $\mu \gg T$ the latter is given by
\begin{equation}
\label{Charge-density}
\rho(\mu)=\frac{|e|\mu^2}{\pi\hbar^2v_F^2}.
\end{equation}

\begin{figure}
\includegraphics[width=0.95\linewidth]{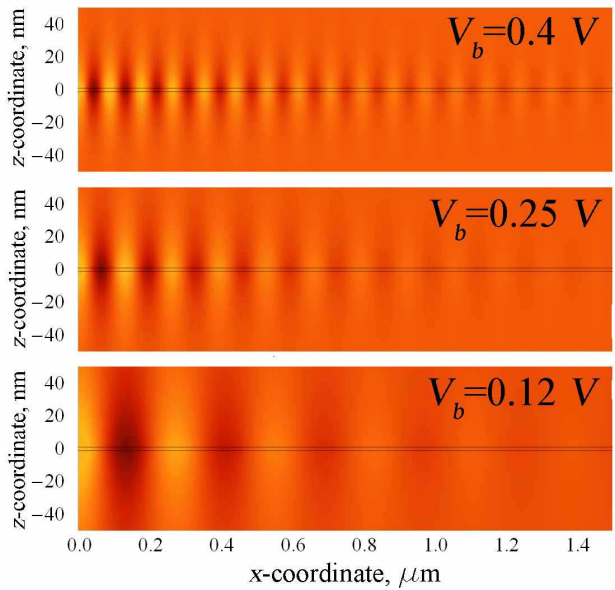}
\caption{\label{Modulation} Illustration of SPP propagation with frequency $\omega/2\pi=25$ THz in 2GL structure ($d=2$ nm) at different bias voltages. The colors correspond to the amplitude $E_x$.}
\end{figure}

The Eqs.~(\ref{Capacitor}--\ref{Charge-density}) allow us to evaluate the damping of SPP of given frequency as a function of bias voltage. The dependence is illustrated in Fig.~\ref{Modulation}: both the propagation length of SPP and ''quality factor'' $\operatorname{Re}k/\operatorname{Im}k$ decrease with decreasing the bias voltage. At almost zero bias the wave damping could be even greater than it is predicted by our equations as the electron-hole scattering will play significant role~\cite{Our-hydrodynamics}.

The plasmon modulators based on 2GL are technologically convenient as they do not require metal gates~\cite{Andersen}. Moreover, the plasmonic spectra in gated graphene structures are linear with velocity much lower than speed of light~\cite{Ryzhii-plasma}, thus coupling of plasmons and electromagnetic radiation could not be efficient. The double layer structures have no such limitations.

\section{Conclusions}
We have obtained the spectra of surface plasmon-polaritons in double-layer graphene structures. The waves can exhibit high ''quality factors'' $\operatorname{Re} k/\operatorname{Im}k$ order of $10^2$ at Fermi energies above 150 meV. The damping of the waves is mostly due to interband transitions, which can be tuned by applying bias voltage between layers. Those interband transitions are strongly indirect, i.e. accounting for spatial dispersion of conductivity is crucial for describing the behavior of SPP. We have shown that spatial dispersion of conductivity shifts the cutoff frequency of SPP to higher values compared to $2\mu/\hbar$ expected from analysis of Pauli blocking.

\section{Acknowledgement}
The work of D.~Svintsov was supported by the grant 14.132.21.1687 of the Russian Ministry of Education and Science. The work of V.~Vyurkov was supported by the grant 11-07-00464 of the Russian Foundation for Basic Research. The work at RIEC was supported by the Japan Science and Technology Agency, CREST and by the Japan Society for Promotion of Science.


\begin{thebibliography}{99}
\bibitem{Asymmetric-Reference}
The asymmetric mode turns out to pass in the region $k>\omega/v_F$ and hence undergoes strong damping due to intraband absorption even at zero ''collision frequency'' $\tau_p^{-1}$.
\end{thebibliography}
\end{document}